\documentclass[12pt,a4paper]{article}

\usepackage{epsfig}
\usepackage{amsmath,amsfonts,amssymb}
\usepackage{t1enc}
\usepackage{cite}
\usepackage{colortbl}
\usepackage{pifont}

\providecommand{\openone}{\leavevmode\hbox{\small1\kern-3.8pt\normalsize1}}

\newcommand{\nn}{\nonumber \\}
\newcommand{\refeq}[1]{Eq.\ (\ref{#1})}

\newcommand{\be}{\begin{equation}}
\newcommand{\ee}{\end{equation}}

\newcommand{\gmu}{\mathcal{G}_\mu}
\newcommand{\G}{\mathcal{G}}

\newcommand{\afb}{A_\text{FB}}
\newcommand{\ac}{A_\text{C}}

\newcommand{\minv}{m_{t\bar{t}}}

\begin{document}

\begin{center}
\begin{Large}
{\bf Shaping the top asymmetry}
\end{Large}

\vspace{0.5cm}
J. A. Aguilar--Saavedra, M. P\'erez-Victoria \\[0.2cm] 
{\it Departamento de F\'{\i}sica Te\'orica y del Cosmos and CAFPE, \\
Universidad de Granada, E-18071 Granada, Spain}
\end{center}

\begin{abstract}
We study different profiles of the distribution of the top forward-backward asymmetry, as a function of the invariant mass of the $t \bar t$ pair. We show that they can be reproduced by one or more light colour octets, while keeping moderate departures of the $t \bar t$ cross section and invariant mass distributions with respect to the Standard Model predictions at Tevatron and LHC.
\end{abstract}

Since their discovery of the top quark~\cite{Abe:1995hr,Abachi:1995iq}, the CDF and D0 collaborations at the Fermilab Tevatron have analysed a large number of events with top-antitop pairs. Interestingly, the measurements of both collaborations in the semileptonic~\cite{Abazov:2007qb,Aaltonen:2008hc,Aaltonen:2011kc} and dileptonic~\cite{CDFAdil} topologies consistently point to an excess in the forward-backward (FB) asymmetry with respect to the Standard Model (SM) predictions~\cite{Kuhn:1998jr,Antunano:2007da,Bernreuther:2010ny,Ahrens:2011uf,Hollik:2011ps}.
Furthermore, in Ref.~\cite{Aaltonen:2011kc}, the CDF Collaboration has reported a nontrivial dependence of the FB asymmetry on the invariant mass $\minv$ of the top-antitop pair in semileptonic events. The following ($t\bar{t}$ rest frame) values were found, at the parton level, in the low and high invariant mass bins:
\begin{align}
& A_{\text{FB},<}\, = -0.116\pm0.153 && (m_{t\bar{t}}<450~\mathrm{GeV}) \,,\nn  
& A_{\text{FB},>}\, = 0.475\pm0.114  && (m_{t\bar{t}}>450~\mathrm{GeV}) \,,
\label{ec:lowhigh}
\end{align}
to be compared with the SM predictions $A_{\text{FB},<}^\text{SM} = 0.040$,
$A_{\text{FB},>}^\text{SM} = 0.088$~\cite{Campbell:1999ah}. 
The separation of the two bins at 450~GeV was chosen to maximise the expected sensitivity at high mass, using as benchmark a heavy colour octet~\cite{Ferrario:2009bz}. A more detailed $\minv$ distribution of the asymmetry was also provided, but only at the data level. 
\begin{figure}[ht]
\begin{center}
\epsfig{file=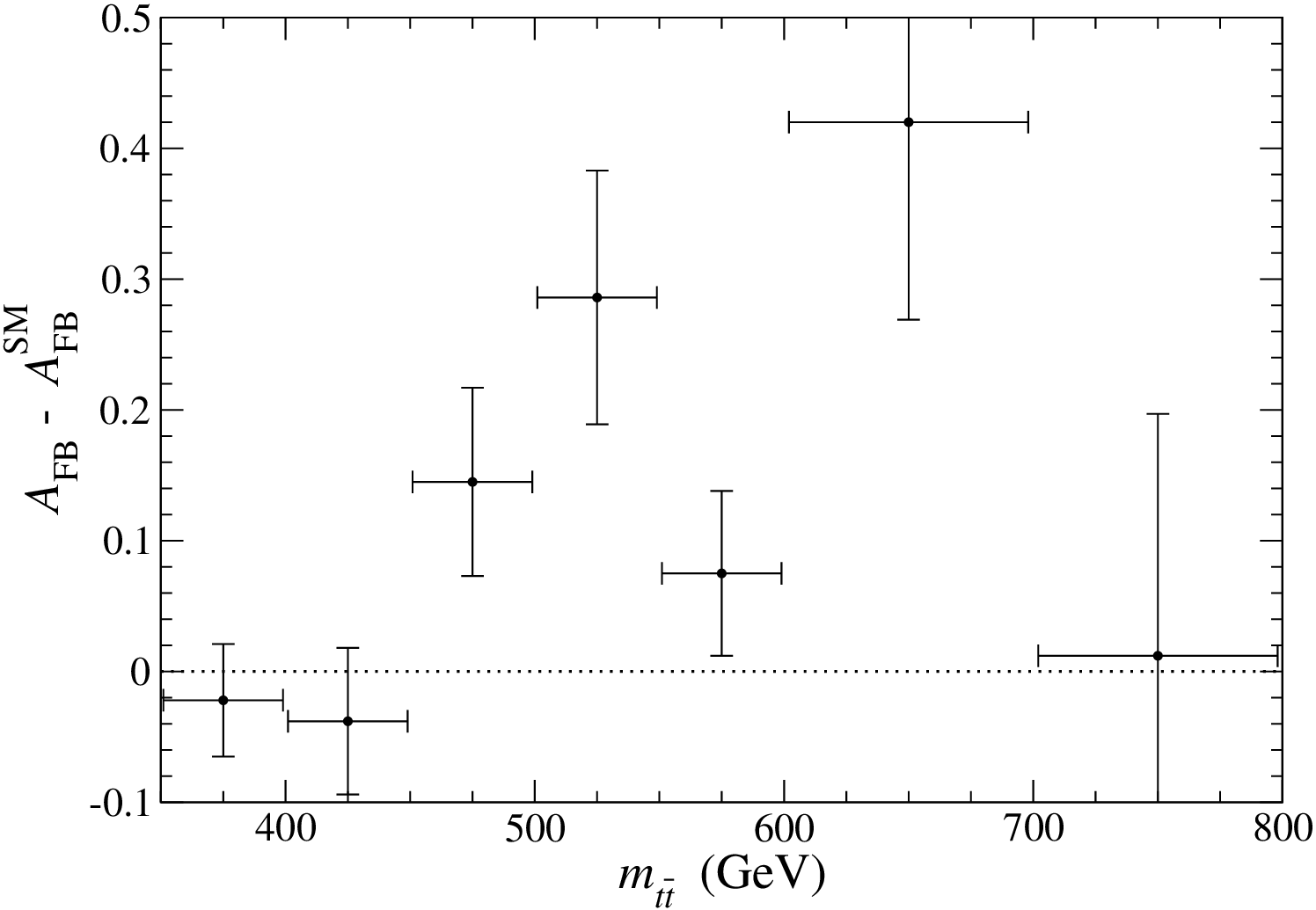,width=10cm} 
\caption{Dependence of the FB asymmetry on $\minv$, from Ref.~\cite{Aaltonen:2011kc}.}
\label{fig:afb-cdf}
\end{center}
\end{figure}
In Fig.~\ref{fig:afb-cdf} we plot the data in that publication, subtracting the SM contribution which is small in all cases. We can observe two clear features beyond the two values in \refeq{ec:lowhigh}: first, the asymmetry in the last bin, $\minv > 700~\mathrm{GeV}$, is much lower than the one in the previous one (actually, it is consistent with zero); second, there is a dip in the $550-600$ GeV bin. Even if, according to Ref.~\cite{Aaltonen:2011kc}, the large statistical errors in the $\minv$ distribution of the asymmetry do not allow any conclusion on the functional dependence, it is quite intriguing that the asymmetries in the two independent samples with positive and negative leptons behave in nearly opposite fashion, as is manifest in Fig.~11 of that reference. That symmetric pattern suggests that the observed distribution is not produced by statistical fluctuations. Furthermore, the full detector simulations performed there to check scenarios with asymmetries beyond the SM indicate that the detector acceptance effects on the asymmetry are largely independent of $\minv$.
It is, nevertheless, way too soon to ascribe this structure to new physics, until these results are either confirmed or refuted by new analyses. In particular, the D0 collaboration has not yet reported a mass-dependent analysis and, besides, the CDF Collaboration has not given the distribution at the parton level. Therefore, it seems sensible at this moment to keep an open mind about the real profile of the FB asymmetry. 

It turns out that almost all the new physics models that have been proposed so far predict roughly the same shape: a FB asymmetry that increases monotonically with the invariant mass in the energy range probed by Tevatron. This behaviour agrees qualitatively with the basic CDF results, although the mass dependence is milder (by at least one sigma) than the one given by the central values in \refeq{ec:lowhigh}, see Ref.~\cite{AguilarSaavedra:2011ug}. In this Letter we consider a scenario in which this situation changes dramatically. We show
that a variety of asymmetry profiles can be generated if the excess in the asymmetry is produced by one or more new colour-octet vector fields exchanged in the $s$ channel, with masses between 300 and 1100~GeV. The presence of relatively light $s$-channel particles is necessary to reproduce non-trivial profiles, and in this case colour octets are required if we want to have interference between the new physics and SM amplitudes. The only two possible $\text{SU}(3)_C\times \text{SU}(2)_L\times \text{U}(1)_Y$ multiplets are~\cite{AguilarSaavedra:2011vw} vector fields $\gmu$ in the $(8,1)_0$ representation, which we call ``gluons'' hereafter, and scalar fields $\Phi$ in the $(8,2)_{-1/2}$, which we will not use in this Letter. Of course, above the $t\bar{t}$ threshold, we need to hide the resonances in the differential cross section. For this, we can resort to large widths, as proposed in Ref.~\cite{Barcelo:2011vk}. In that work, the large widths are achieved by opening new decay channels of the gluons into additional new particles. Here, we take a phenomenological approach and adjust the widths freely, since our basic results (related to $t\bar{t}$ production and not to other possible collider signals of extra particles) are quite insensitive to the particular mechanism that makes the gluon resonances broad. In any case, we will later discuss different options to enhance the widths.

Generating the FB asymmetry with light particles in the $s$ channel has an important bonus~\cite{Barcelo:2011vk,Barcelo:2011fw}. Most explanations of the FB asymmetry in terms of new physics predict an increase of $t\bar{t}$ production in the tail of the invariant mass distribution~\cite{AguilarSaavedra:2011vw}. However, no such effect has been observed in the recent LHC data~\cite{CMStail}. This leads to strong constraints on the available parameter space of the different models~\cite{AguilarSaavedra:2011hz}. If the agreement with the SM persits with the increasing precision, even tighter bounds will be imposed and some scenarios will be completely ruled out~\cite{AguilarSaavedra:2011ug}. In this situation, it is crucial to study models that do not enhance the $t\bar{t}$ tail. Our light gluons have the virtue of not producing large deviations in the cross section at energies far above their masses, as we will show. Thus, they comply with these LHC constraints.

We will study the asymmetry produced by one, two or three light gluons $\G_i$ of masses $M_i$. The subscript $i$ indicates the specific gluon. The relevant interactions are given by
\be
\mathcal{L}_\mathrm{int} = - \sum_i  \left(-g^q_i \bar{q}_L\gamma_\mu \frac{\lambda^a}{2} q_L 
+ g^q_i \bar{u}_R \gamma_\mu \frac{\lambda^a}{2} u_R 
+ g^q_i \bar{d}_R \gamma_\mu \frac{\lambda^a}{2} d_R + g^t_i \bar{t}_R \gamma_\mu \frac{\lambda^a}{2} t_R \right) \G_i^\mu  \, ,
\ee
where $q_L = (u,d)^T$ is the light-quark doublet. The couplings to the light quarks are chosen to be axial, so that the interference with the SM amplitude in the total cross section vanishes. Having also axial couplings to the top quark would then maximise the asymmetry, relative to the increase in the total cross section $\Delta \sigma$. Still, we have chosen chiral couplings to the right-handed top quark, in order to avoid problems from flavour-changing neutral currents~\cite{Bai:2011ed} and possibly $Z\to b\bar{b}$.  This notwithstanding, we will also show the results for a completely axial gluon in one particular example below. 

The impact of each gluon  $\G_i$ on the amplitude (and hence on the generated asymmetry) is proportional to the product of couplings to light and top quarks, $X_i= g^q_i g^t_i$. The widths of the gluons, on the other hand, depend on the separate couplings and also on possible additional decay modes. For definiteness, we  use for each gluon an energy-dependent width with the same functional form as the one induced by the decay into right-handed tops, and an intensity controlled by an independent parameter $r_i$, such that $M_i \Gamma_i = r_i \gamma s$, with
\be
\gamma= \frac{1}{48\pi}  \left(1-\frac{m_t^2}{\hat s}\right) \left(1-4\frac{m_t^2}{s}\right)^{\frac{1}{2}} \theta(s-4m_t^2) \,,
\ee
$s$ being the partonic centre of mass energy.
In the particular case when $t\bar{t}$ is the dominant decay channel, $r_i\simeq (g_i^t)^2$.
In addition, for widths comparable to the mass splittings, it is necessary to take into account the mixing of widths induced by the common decay channels, see~\cite{Cacciapaglia:2009ic} and references therein. This involves inverting the two-point function at the one-loop level. The amplitude is proportional to an ``effective propagator'' $P_\mathrm{eff}^{\mu\nu}=\sum_{i,j} g^q_i g^t_j \Delta_{ij}^{\mu\nu}$, where $\Delta_{ij}^{\mu\nu}$ is the Feynman propagator from a gluon $\G_i^\mu$ to a gluon $\G_j^\nu$. For three gluons, we find
\begin{equation}
P_\mathrm{eff}^{\mu\nu}=\eta^{\mu\nu} \; \frac{\mathcal{N}}{\mathcal{D}} \,,
\end{equation}
with
\begin{eqnarray}
\mathcal{N} & = & X_1 (s-M_2^2)(s-M_3^2) + X_2 (s-M_1)^2(s-M_3^2)+X_3 (s-M_1^2)(s-M_2^2) \,, \notag \\
\mathcal{D} & = & (s-M_1)^2 (s-M_2^2) (s-M_3^2) + i \gamma s \left[ r_1 (s-M_2^2)(s-M_3^2) \right. \notag \\
& & \left. + r_2 (s-M_1)^2(s-M_3^2)+r_3 (s-M_1^2)(s-M_2^2) \right] \,.
\end{eqnarray}
We have just written the relevant part, proportional to the metric. The longitudinal part can be neglected because of the small mass of the $u$ and $d$ quarks.
In the case of one or two gluons, the same formula is valid, just setting the $X_i$ and $r_i$ of the non-active gluons to zero. The cross sections and asymmetries are calculated incorporating the matrix elements in the leading-order generator {\tt Protos}~\cite{AguilarSaavedra:2008gt}.

We are ready to study explicit examples. We consider six benchmarks, designed to give distinctively different asymmetry shapes. This should be sufficient to illustrate the possibilities of our scenario. The first three models have only one gluon, the next two contain two gluons, and the last one has three. The couplings, masses and width parameters in all these models are collected in Table~\ref{tab:models}, together with their predictions for the new physics contributions to the asymmetries in the low- and high-mass bins. The coupling to the light quarks is constrained by dijet data at Tevatron~\cite{Aaltonen:2008dn} and LHC~\cite{Aad:2011aj}. We include in Table~\ref{tab:models} the maximum $g_i^q$ consistent with constraints, as well as the corresponding minimum $g_i^t$ for the given values of $X_i$. Note
that, because the coupling to the top quark is larger than for light quarks, for gluon masses larger than $2m_t$ a sizable branching ratio is invisible in dijet final states. Moreover, for the heavier masses $M_i=870, 1050$ GeV the gluons are rather wide, so that these dijet constraints are conservative and even larger couplings to the light quarks (and smaller couplings to the top) would be allowed.\footnote{For example, the analysis in Ref.~\cite{Aad:2011aj} cuts on a window of $0.3 M$ around the gluon mass, which is narrower than the intrinsic width of the $M=870,1050$ GeV gluons, in order to obtain the limits.} In any case, the numbers presented in the table show that the different asymmetry profiles can be reproduced with moderate couplings to the top quark and an extra enhancement of the width by decays to other final states.

\begin{figure}[htb]
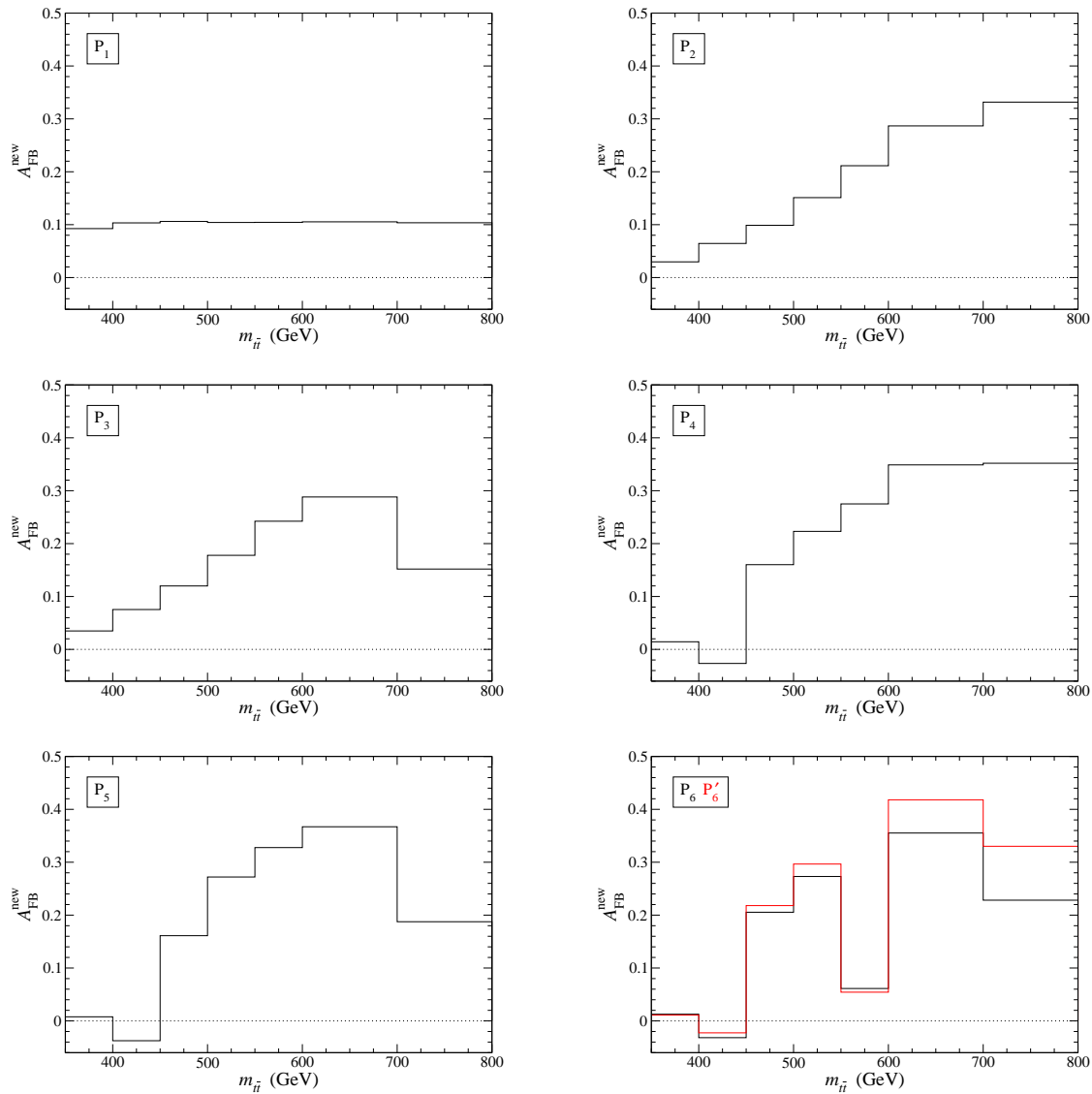

\begin{center}
\begin{tabular}{ccc}
\epsfig{file=Figs/afb-P1.eps,width=6.7cm,clip=} & \quad\quad &
\epsfig{file=Figs/afb-P2.eps,width=6.7cm,clip=} \\[2mm]
\epsfig{file=Figs/afb-P3.eps,width=6.7cm,clip=} & &
\epsfig{file=Figs/afb-P4.eps,width=6.7cm,clip=} \\[2mm]
\epsfig{file=Figs/afb-P5.eps,width=6.7cm,clip=} & &
\epsfig{file=Figs/afb-P6.eps,width=6.7cm,clip=}
\end{tabular}
\caption{FB asymmetry in bins of $t \bar t$ invariant mass.}
\label{fig:shape}
\end{center}
\end{figure}

\begin{figure}[htb]
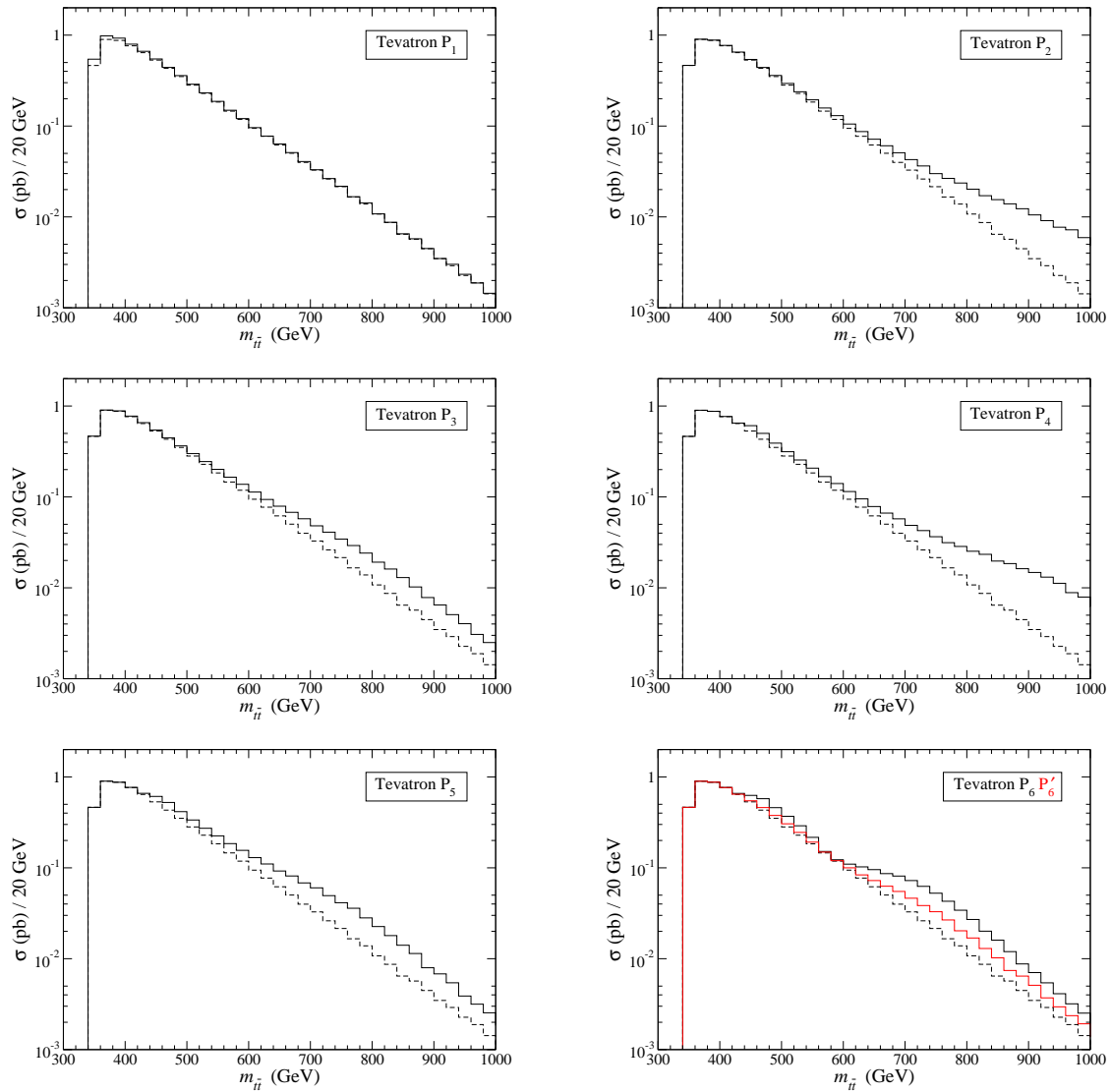

\begin{center}
\begin{tabular}{ccc}
\epsfig{file=Figs/mtt-t-P1.eps,width=6.7cm,clip=} & \quad\quad &
\epsfig{file=Figs/mtt-t-P2.eps,width=6.7cm,clip=} \\[2mm]
\epsfig{file=Figs/mtt-t-P3.eps,width=6.7cm,clip=} & &
\epsfig{file=Figs/mtt-t-P4.eps,width=6.7cm,clip=} \\[2mm]
\epsfig{file=Figs/mtt-t-P5.eps,width=6.7cm,clip=} & &
\epsfig{file=Figs/mtt-t-P6.eps,width=6.7cm,clip=}
\end{tabular}
\caption{Invariant mass distribution at Tevatron for the SM (dashed lines) and the six reference models(solid).}
\label{fig:mtt-tev}
\end{center}
\end{figure}

\begin{figure}[htb]
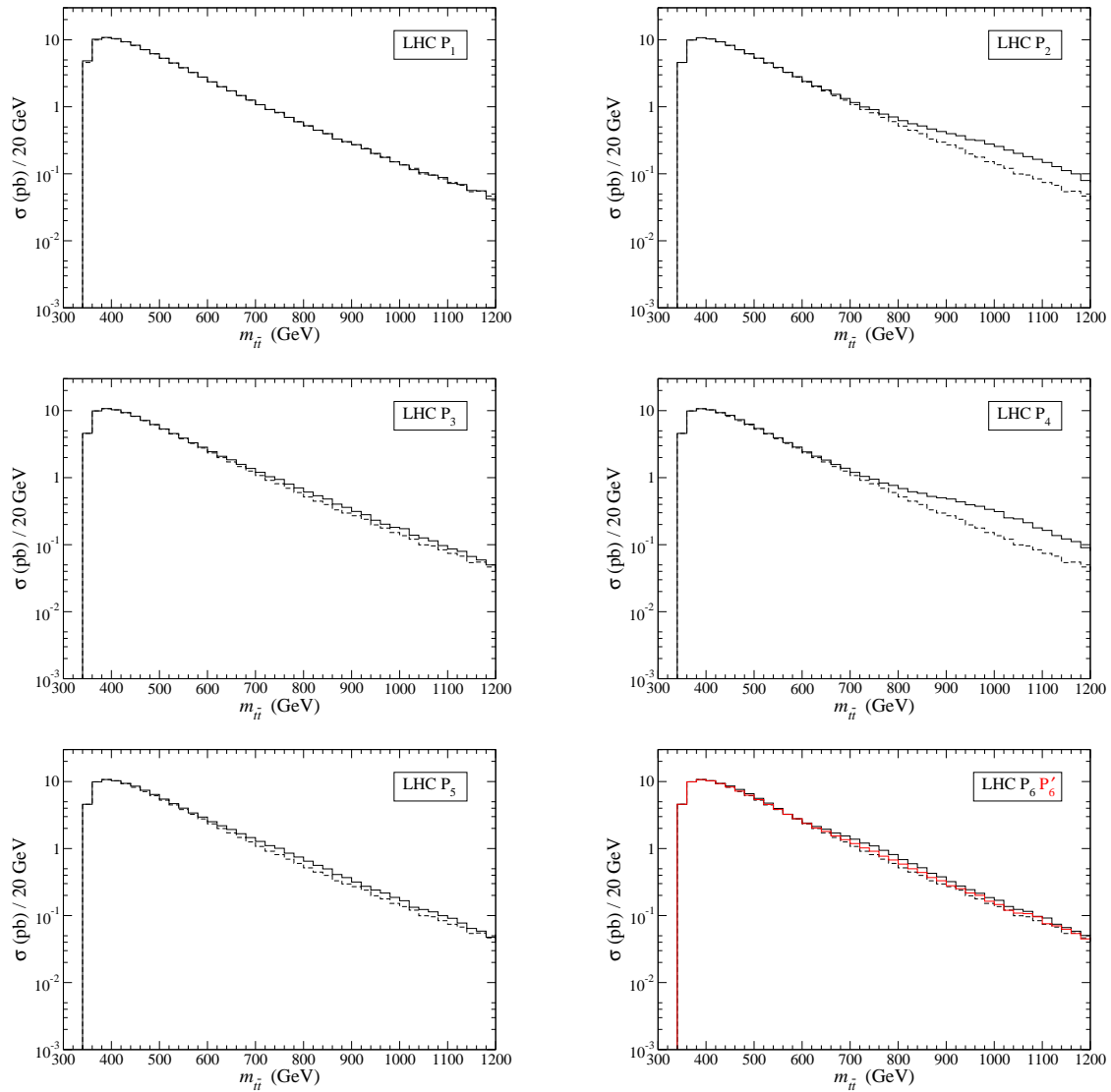

\begin{center}
\begin{tabular}{ccc}
\epsfig{file=Figs/mtt-l-P1.eps,width=6.7cm,clip=} & \quad\quad &
\epsfig{file=Figs/mtt-l-P2.eps,width=6.7cm,clip=} \\[2mm]
\epsfig{file=Figs/mtt-l-P3.eps,width=6.7cm,clip=} & &
\epsfig{file=Figs/mtt-l-P4.eps,width=6.7cm,clip=} \\[2mm]
\epsfig{file=Figs/mtt-l-P5.eps,width=6.7cm,clip=} & &
\epsfig{file=Figs/mtt-l-P6.eps,width=6.7cm,clip=}
\end{tabular}
\caption{Invariant mass distribution at LHC for the SM (dashed lines) and the six reference models (solid).}
\label{fig:mtt-lhc}
\end{center}
\end{figure}

\begin{table}[htb]
\begin{center}
\begin{tabular}{cccccccc}
Model & $M_i$ & $X_i$ & $|g_i^q|_\text{max}$ & $|g_i^t|_\text{min}$ & $r_i$ & $A_{\text{FB},<}^\text{new}$ & $A_{\text{FB},>}^\text{new}$ \\
\hline
$\text{P}_1$ & 320 GeV  &  0.224  & 0.23 & 0.96 & --  & 0.096  & 0.105 \\[2mm]
$\text{P}_2$ & 1050 GeV & -1.6    & 0.80 & 2.0  & 64  & 0.045  & 0.178 \\[2mm]
$\text{P}_3$ & 870 GeV  & -1.2    & 0.57 & 2.1  & 100 & 0.052  & 0.180 \\[3mm]
$\text{P}_4$ & \begin{tabular}{c} 450 GeV \\ 1050 GeV \end{tabular} 
             & \begin{tabular}{c} 0.0644 \\ -1.84 \end{tabular}
             & \begin{tabular}{c} 0.20 \\ 0.83 \end{tabular}
             & \begin{tabular}{c} 0.33 \\ 2.2 \end{tabular}
             & \begin{tabular}{c} 16 \\ 64 \end{tabular}
             & -0.004 & 0.238  \\[5mm]
$\text{P}_5$ & \begin{tabular}{c} 450 GeV \\ 870 GeV \end{tabular}
             & \begin{tabular}{c} 0.0975 \\ -1.3 \end{tabular}
             & \begin{tabular}{c} 0.21 \\ 0.59 \end{tabular}
             & \begin{tabular}{c} 0.46 \\ 2.2 \end{tabular}
             & \begin{tabular}{c} 25 \\ 100 \end{tabular}
             & -0.014 & 0.243 \\[5mm]
$\text{P}_6$ & \begin{tabular}{c} 450 GeV \\ 570 GeV \\ 870 GeV \end{tabular}
             & \begin{tabular}{c} 0.105 \\ -0.049 \\ -1.4 \end{tabular}
             & \begin{tabular}{c} 0.22 \\ 0.17 \\ 0.60 \end{tabular}
             & \begin{tabular}{c} 0.49 \\ 0.28 \\ 2.3 \end{tabular}
             & \begin{tabular}{c} 25 \\ 25 \\ 100 \end{tabular}
             & -0.006 & 0.227 
\end{tabular}
\caption{Parameters used for the six models representative of the different profiles, and new physics contributions to the asymmetries in the low- and high-mass bins.}
\label{tab:models}
\end{center}
\end{table}
We show in Fig.~\ref{fig:shape} the resulting distributions of the FB asymmetry as a function of $\minv$. 
We see that, as promised, quite diverse $\minv$ profiles are generated. The corresponding distributions of the total cross section are shown, together with the SM one, in Fig.~\ref{fig:mtt-tev}  for Tevatron and in Fig.~\ref{fig:mtt-lhc} for LHC.

In all cases, we have adjusted the overall size of the couplings to give a new physics contribution to the inclusive asymmetry $\afb^\text{new} = 0.1$, as resulting from the CDF measurement. Minor details of the distributions, such as the sign of the asymmetry in the first two bins, and the amount of the decrease in the last one, can be tailored by a suitable choice of the masses and couplings of the gluons. In any case, the selected examples are representative of typical behaviours. Let us comment in turn on the specific features of each of the models.
 
\underline{Model $\text{P}_1$} gives a flat asymmetry profile. To achieve this, we have extended the SM with just one gluon of mass $M=320\, \mathrm{GeV}$, below but sufficiently far from the $t\bar{t}$ production threshold. The couplings of the top and the light quarks are chosen to have the same sign, so that $X$ is positive and thus a positive asymmetry is generated at $\minv$ greater than $M$.\footnote{This is also the case in the model of Ref.~\cite{Tavares:2011zg}, which appeared as we were finishing the writing of the present work. There, the top asymmetry is explained by a light axigluon of mass $M \sim 420$ GeV and universal couplings to all the quarks. The value of the mass is chosen to give a change of sign in the asymmetry at the interference level, to mimic \refeq{ec:lowhigh}. At variance with our model $\text{P}_1$, new particles are required to dilute the resonance in the cross section.}
 This scenario has several advantages. First, because top pairs cannot be produced resonantly, the quadratic new physics term never dominates, and the cross section distribution follows precisely the one of the SM, as it can be seen in Figs.~\ref{fig:mtt-tev} and~\ref{fig:mtt-lhc}. In particular, we do not need to enhance the width. Another important feature is that the couplings can be rather weak, which renders the scenario quite robust under future dijet and flavour constraints. Of course, a flat profile does not agree with the results in \refeq{ec:lowhigh}, but by selecting lower masses $M$ the asymmetry can have a mild increase with $\minv$.
And, as we have remarked above, these shapes still need more statistics and independent confirmation from the D0 collaboration. 

\underline{Model $\text{P}_2$} gives an asymmetry profile that increases in all the Tevatron $\minv$ range. For this, we use a relatively heavy gluon of mass $M = 1050$ GeV. Observe that, with these parameters, the gluon produces an excess in the cross section at high invariant masses that is in some tension with the Tevatron measurement~\cite{Aaltonen:2009iz}, since the cross section at the last measured bin $\minv \geq 800$ GeV would be at $2\sigma$ from data. The enhancement would also be visible at the LHC with increased precision, with a tail $\sigma(\minv > 1~\text{TeV})$ of 1.8 times the SM cross section. We note that this asymmetry shape, which roughly agrees with \refeq{ec:lowhigh}, is similar to the one obtained in models with new particles in the $t$ and $u$ channel or with heavy new physics. However, as we will see below, the prediction for charge asymmetries at LHC are very different.

\underline{Model $\text{P}_3$} produces a rising asymmetry that decreases above 700 GeV. The mass of the gluon in this case is 870 GeV (similar to the one in Refs.~\cite{Barcelo:2011vk,Barcelo:2011fw}). The increase in the cross section is small at Tevatron and very small at LHC, where the presence of the gluon would be invisible.

\underline{Model $\text{P}_4$} gives a profile similar to the one in model $\text{P}_2$, but with much smaller asymmetries in the first two bins (negative in the second one), which improve the agreement with \refeq{ec:lowhigh}. To achieve this, we add a new gluon to the one in model $\text{P}_2$, with mass $M_2 = 450$ GeV and same-sign (smaller) light and top quark couplings to decrease the asymmetry below 450 GeV.
We have checked that the shape is similar to the one in Ref.~\cite{Tavares:2011zg}, with the difference that in our model the asymmetries in the first two bins are small, while in the model of~\cite{Tavares:2011zg} the first bin has a large, negative asymmetry. The cross section at high invariant masses is in tension with the measurements, as in model $\text{P}_2$, due to the high mass of the heavier gluon $M_1 = 1050$ GeV.

\underline{Model $\text{P}_5$} is similar to model $\text{P}_3$, but with small asymmetry in the first two bins, negative in the second one. It contains the same gluon of model $\text{P}_3$ plus a lighter one with $M_2 = 450$ GeV, with same-sign quark couplings, which allows for a better agreement with~\refeq{ec:lowhigh}.
The departures in the cross section are quite small, both for Tevatron and LHC. 

\underline{Model $\text{P}_6$} produces a camel-like profile that resembles the one in Fig.~\ref{fig:afb-cdf}. To accomplish this, we need three gluons. The first two have masses as in model $\text{P}_5$, while the fourth one is located in between, at $M_3 = 570$ GeV. While the shape of the asymmetry is interesting, too large an excess is produced in the cross section at Tevatron. For these reasons, we have also studied model $\text{P}_6'$ (red solid lines), in which the couplings of the three gluons are axial for both the light and the top quark. In this case, the couplings to the top quark are divided by two to have similar widths as in model $\text{P}_6$ ($r_i = 100,\;29 ,\;26$ for $M_i = 870,\;450,\;570$ GeV, respectively), and the couplings to light quarks are slightly adjusted ($X_i = -0.65,\;0.049,\;-0.023$) to keep an inclusive asymmetry $\afb^\text{new} = 0.1$. Then, the cross section is reduced relative to the asymmetry, and it deviates little from the SM one, both at Tevatron and LHC. Notice that the shapes differ slightly because the widths are not exactly equal in the axial case. On the other hand, axigluons that couple in a non-universal way may induce flavour changing neutral currents, and are thus subject to additional constraints. As we discuss below, they are satisfied by the axigluons in model $\text{P}_6'$.
 Let us also point out that this setup with several overlapping broad 
resonances in the same channel is reminiscent of unparticle physics~\cite{Georgi:2007ek}, as discussed in Ref.~\cite{PerezVictoria:2008pd}. However, usual unparticle theories (see Ref.~\cite{Chen:2010hm} for an application to the top asymmetry) do not give couplings with different sign at different energies.

All our benchmarks except $\text{P}_1$ require large widths, especially for the heaviest gluon $\G_1$. The simplest way to generate them is to use the top quark itself, and enhance its contribution by choosing large couplings $g^t_i$. For instance, if $g^t_1$ is between 8 and 10, the necessary $r_1$ from 64 to 100 is generated. A large hierarchy $g^t_i \gg g_i^q$ is natural in models with extra dimensions~\cite{Barcelo:2011fw,Djouadi:2009nb,Delaunay:2010dw,Alvarez:2010js}, and an extra advantage of large top couplings is that the couplings to the light quarks can be small, thus evading limits from measurements of dijet cross sections.
However, such strong couplings would give rise to important radiative corrections to our tree-level results, and could even drive the theory into the nonperturbative regime. Therefore, a different mechanism may be required. Let us briefly comment on some alternatives. 
First, we can just turn on a coupling to the right-handed $b$ quark to open a new decay mode. Flavour changing neutral currents can be avoided by a convenient alignment of the right-handed quarks. On the other hand, too large a $b$ coupling would give rise to an excess of $b \bar b$ dijets.
Second, we can consider completely axial gluons, as in model $\text{P}_6'$, to improve the $\afb^\text{new} / \Delta \sigma$ ratio, so that effects on the invariant mass distribution are less significant and smaller widths are needed. Axigluons also help to generate larger widths, as the $t_{R,L}$, $b_{R,L}$ channels are open (see also the recent Ref.~\cite{Alvarez:2011hi}). An important problem is, however, that in this case there are unavoidable flavour bounds from a combination of data in neutral $B$, $K$ and $D$ meson mixing~\cite{Bai:2011ed}. This prevents the coupling to the $t$ and $b$ quarks from being too large. 
 In the case of our axial model $\text{P}_6'$, for instance, we can neglect the contribution of the two lighter gluons, which have small $|g^t_i-g^q_i|$. Then, these flavour limits require that the couplings of the heavier gluon fulfill $|g^t_1-g^q_1|\lesssim 6.4$. Choosing for example $g^t_1=5$, $g_1^q=-0.13$,  we comply with both flavour and dijet constraints,\footnote{For this model we have $|g_i^q|_\text{max} = 0.59,\;0.22,\;0.17$ from dijet constraints, implying $|g_i^t|_\text{min} = 1.1,\;0.23,\;0.13$.}
and we get the required $X_1=-0.65$ and $r_1=100$. Note also that the virtual effects of the axigluons on the $Z\bar{b}b$ coupling are proportional to $m_b^2$ and can be neglected~\cite{Chivukula:2010fk}.
Finally, we can invoke additional new particles to increase the width, as in the stealth gluon proposal in Ref.~\cite{Barcelo:2011vk} (see also Refs.~\cite{Bai:2010dj,Bai:2011ed,Tavares:2011zg}). The corresponding limits and signals depend on the specific scenario. Let us just point out, in this regard, that it is possible that the new particles lie between two of our gluons, with the consequence that only the width of the heavier gluon would be increased. One could even conceive a scenario in which the heavier gluons decay into the lighter ones. 

\begin{figure}[t]
\begin{center}
\epsfig{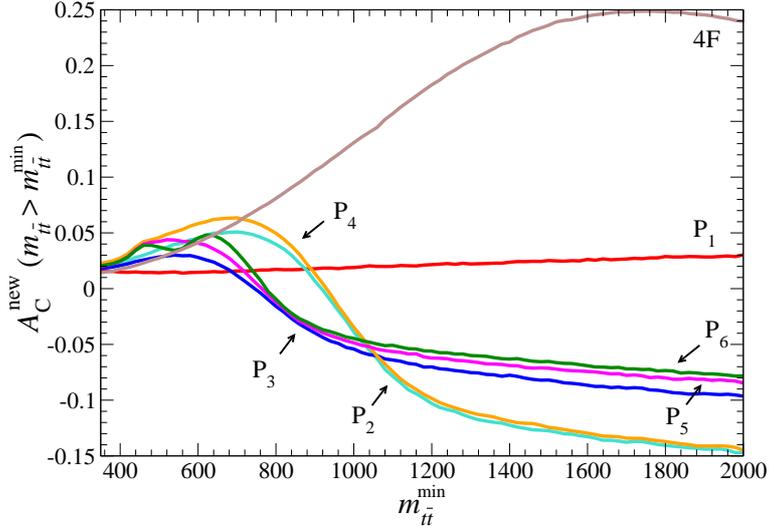} 
\caption{New physics contributions to the charge asymmetry at LHC in the different models as a function of the lower cut $\minv^\mathrm{min}$.}
\label{fig:afb-mtt}
\end{center}
\end{figure}
Finally, it is interesting, in view of the upcoming measurements of charge asymmetries at LHC, that this scenario with light gluons leads to predictions that are strikingly different from the ones in simple models~\cite{AguilarSaavedra:2011ug} when the events with large $t\bar{t}$ invariant mass are selected. To show this, we plot in Fig.~\ref{fig:afb-mtt} the charge asymmetry $\ac^\text{new}$ predicted by our benchmarks for events with $\minv$ larger than a varying cutoff $\minv^\mathrm{min}$. We use the following definition of the charge asymmetry:
\begin{equation}
\ac = \frac{N(\Delta > 0) - N(\Delta < 0)}{N(\Delta > 0) + N(\Delta < 0)} \,,
\label{ec:ACMS}
\end{equation}
with $\Delta = |\eta_t|- |\eta_{\bar t}|$, as used by the CMS Collaboration~\cite{afbCMS}. Here, $\eta$ stand for the pseudo-rapidities in the laboratory frame, and $N$ for the number of $t \bar t$ events. 
(Using the rapidity difference $\Delta = |y_t|- |y_{\bar t}|$ leads to the same numerical asymmetry, as it is also the case for a FB asymmetry defined by taking the forward direction as the one of the longitudinal boost of the $t \bar t$ system~\cite{Krohn:2011tw}). 
For comparison, we also include in Fig.~\ref{fig:afb-mtt} the prediction for a very heavy axigluon, described by four-fermion (4F) operators, with $C/\Lambda^2 = 0.66~\text{TeV}^{-2}$~\cite{AguilarSaavedra:2011hz}. We see that the heavy gluon predicts a positive charge asymmetry that increases with $\minv^\mathrm{min}$. In contrast, models $\text{P}_{3,5,6}$ ($\text{P}_{2,4}$) predict a charge asymmetry that becomes negative at $\minv^\mathrm{min}\approx700$ (900) GeV. Model $\text{P}_1$ lies roughly in the middle, and gives a small positive charge asymmetry for all values of $\minv^\mathrm{min}$.

To conclude, we have shown how models with one or more light gluons can give rise to different shapes for the $\minv$ distribution of the FB asymmetry. In particular, these models can reproduce different features of the CDF data in Ref.~\cite{Aaltonen:2011kc}, including the camel-like profile in Fig.~\ref{fig:afb-cdf}. In general, the gluons must have large widths in order not to distort too much the $\minv$ distributions of the cross sections at Tevatron and LHC. We have discussed different mechanisms that can give rise to widths of that size. To decide which mechanisms are viable, a more detailed analysis of the constraints is necessary, but this is beyond the scope of the present work. We have also shown the predictions of these models for the charge asymmetry at LHC. In addition, we have proposed a model with a gluon with mass below the $t\bar{t}$ threshold, and the same signs for the top and light quark couplings. This model does not alter the SM $t \bar t$ cross sections and distributions. On the other hand, it predicts a flat profile (or smoothly increasing with $\minv$, for lower gluon masses) for both the FB and charge asymmetries. Some of the profiles we have shown are disfavoured by the mass-dependent findings of the CDF Collaboration~\cite{Aaltonen:2011kc}. However, we still have to see what the D0 and LHC experiments have to say about the mass dependence of the FB asymmetry. 

{\it Note added.} After the submission of this Letter the new measurement by the D0 Collaboration was made public~\cite{Collaboration:2011rq}. An unfolded measurement of the mass dependence has not yet been presented. At the reconstruction level, the asymmetry does not exhibit a statistically significant enhancement at high invariant masses. This mild dependence would correspond to the profile $\text{P}_1$ in our classification, achievable with a light gluon before the $t \bar t$ threshold. This profile is also in agreement with the mass-dependence (at the reconstructed level) in the new CMS measurements~\cite{afbCMS2}. The rest of models, with a growing asymmetry as corresponds to the CDF result~\cite{Aaltonen:2011kc}, are disfavoured by those other two measurements.

\section*{Acknowledgements}

We thank Manel Masip and Jos\'e Santiago for discussions. This work has been partially supported by projects FPA2010-17915 (MICINN), FQM 101 and FQM 437 (Junta de Andaluc\'{\i}a) and CERN/FP/116397/2010 (FCT).


\begin{thebibliography}{99}

\bibitem{Abe:1995hr}
  F.~Abe {\it et al.}  [CDF Collaboration],
  Phys.\ Rev.\ Lett.\  {\bf 74}, 2626 (1995)
  [arXiv:hep-ex/9503002].

\bibitem{Abachi:1995iq}
  S.~Abachi {\it et al.}  [D0 Collaboration],
  Phys.\ Rev.\ Lett.\  {\bf 74}, 2632 (1995)
  [arXiv:hep-ex/9503003].

\bibitem{Abazov:2007qb}
  V.~M.~Abazov {\it et al.}  [D0 Collaboration],
  Phys.\ Rev.\ Lett.\  {\bf 100}, 142002 (2008)
  [arXiv:0712.0851 [hep-ex]].

\bibitem{Aaltonen:2008hc}
T.~Aaltonen {\it et al.}  [CDF Collaboration],
  Phys.\ Rev.\ Lett.\  {\bf 101}, 202001 (2008)
  [arXiv:0806.2472 [hep-ex]].

\bibitem{Aaltonen:2011kc}
T.~Aaltonen {\it et al.}  [CDF Collaboration],
  Phys.\ Rev.\  D {\bf 83}, 112003 (2011)
  [arXiv:1101.0034 [hep-ex]].

\bibitem{CDFAdil}
T.~Aaltonen {\it et al.} [CDF Collaboration], CDF note 10436.

\bibitem{Kuhn:1998jr}
  J.~H.~Kuhn and G.~Rodrigo,
  Phys.\ Rev.\ Lett.\  {\bf 81}, 49 (1998)
  [arXiv:hep-ph/9802268].

\bibitem{Antunano:2007da}
  O.~Antunano, J.~H.~Kuhn and G.~Rodrigo,
  Phys.\ Rev.\  D {\bf 77}, 014003 (2008)
  [arXiv:0709.1652 [hep-ph]].

\bibitem{Bernreuther:2010ny}
  W.~Bernreuther and Z.~G.~Si,
  Nucl.\ Phys.\  B {\bf 837}, 90 (2010)
  [arXiv:1003.3926 [hep-ph]].

\bibitem{Ahrens:2011uf}
  V.~Ahrens, A.~Ferroglia, M.~Neubert, B.~D.~Pecjak and L.~L.~Yang,
  arXiv:1106.6051 [hep-ph].

\bibitem{Hollik:2011ps}
  W.~Hollik and D.~Pagani,
  arXiv:1107.2606 [hep-ph].

\bibitem{Campbell:1999ah}
  J.~M.~Campbell and R.~K.~Ellis,
  Phys.\ Rev.\  D {\bf 60}, 113006 (1999)
  [arXiv:hep-ph/9905386].

\bibitem{Ferrario:2009bz}
  P.~Ferrario and G.~Rodrigo,
  Phys.\ Rev.\  D {\bf 80}, 051701 (2009)
  [arXiv:0906.5541 [hep-ph]].

\bibitem{AguilarSaavedra:2011ug}
  J.~A.~Aguilar-Saavedra and M.~P\'erez-Victoria,
  arXiv:1107.0841 [hep-ph].

\bibitem{AguilarSaavedra:2011vw}
  J.~A.~Aguilar-Saavedra and M.~P\'erez-Victoria,
  JHEP {\bf 1105}, 034 (2011)
  [arXiv:1103.2765 [hep-ph]].

\bibitem{Barcelo:2011vk}
  R.~Barcelo, A.~Carmona, M.~Masip and J.~Santiago,
  arXiv:1106.4054 [hep-ph].

\bibitem{Barcelo:2011fw}
  R.~Barcelo, A.~Carmona, M.~Masip and J.~Santiago,
  Phys.\ Rev.\  D {\bf 84}, 014024 (2011)
  [arXiv:1105.3333 [hep-ph]].

\bibitem{CMStail}
CMS Collaboration, note CMS PAS EXO-11-055.

\bibitem{AguilarSaavedra:2011hz}
  J.~A.~Aguilar-Saavedra and M.~P\'erez-Victoria,
  arXiv:1105.4606 [hep-ph].

\bibitem{Bai:2011ed}
  Y.~Bai, J.~L.~Hewett, J.~Kaplan and T.~G.~Rizzo,
  JHEP {\bf 1103}, 003 (2011)
  [arXiv:1101.5203 [hep-ph]].

\bibitem{Cacciapaglia:2009ic}
  G.~Cacciapaglia, A.~Deandrea, S.~De Curtis,
  Phys.\ Lett.\  {\bf B682 } (2009)  43-49.
  [arXiv:0906.3417 [hep-ph]].

\bibitem{AguilarSaavedra:2008gt}
  J.~A.~Aguilar-Saavedra,
  Nucl.\ Phys.\  B {\bf 804}, 160 (2008)
  [arXiv:0803.3810 [hep-ph]].

\bibitem{Aaltonen:2008dn}
  T.~Aaltonen {\it et al.}  [CDF Collaboration],
  Phys.\ Rev.\  D {\bf 79}, 112002 (2009)
  [arXiv:0812.4036 [hep-ex]].

\bibitem{Aad:2011aj}
  G.~Aad {\it et al.}  [ATLAS Collaboration],
  New J.\ Phys.\  {\bf 13} (2011) 053044
  [arXiv:1103.3864 [hep-ex]].

\bibitem{Tavares:2011zg}
  G.~M.~Tavares and M.~Schmaltz,
  arXiv:1107.0978 [hep-ph].

\bibitem{Aaltonen:2009iz}
  T.~Aaltonen {\it et al.}  [CDF Collaboration],
  Phys.\ Rev.\ Lett.\  {\bf 102}, 222003 (2009)
  [arXiv:0903.2850 [hep-ex]].

\bibitem{Georgi:2007ek}
  H.~Georgi,
  Phys.\ Rev.\ Lett.\  {\bf 98}, 221601 (2007)
  [arXiv:hep-ph/0703260].

\bibitem{PerezVictoria:2008pd}
  M.~P\'erez-Victoria,
  JHEP {\bf 0901}, 011 (2009)
  [arXiv:0808.4075 [hep-ph]].

\bibitem{Chen:2010hm}
  C.~H.~Chen, G.~Cvetic and C.~S.~Kim,
  Phys.\ Lett.\  B {\bf 694}, 393 (2011)
  [arXiv:1009.4165 [hep-ph]].


\bibitem{Djouadi:2009nb}
  A.~Djouadi, G.~Moreau, F.~Richard and R.~K.~Singh,
  Phys.\ Rev.\  D {\bf 82}, 071702 (2010)
  [arXiv:0906.0604 [hep-ph]];
  A.~Djouadi, G.~Moreau and F.~Richard,
  Phys.\ Lett.\  B {\bf 701}, 458 (2011)
  [arXiv:1105.3158 [hep-ph]].

\bibitem{Delaunay:2010dw}
  C.~Delaunay, O.~Gedalia, S.~J.~Lee, G.~Perez and E.~Ponton,
  Phys.\ Rev.\  D {\bf 83}, 115003 (2011)
  [arXiv:1007.0243 [hep-ph]];
  arXiv:1101.2902 [hep-ph].

\bibitem{Alvarez:2010js}
  E.~Alvarez, L.~Da Rold and A.~Szynkman,
  JHEP {\bf 1105}, 070 (2011)
  [arXiv:1011.6557 [hep-ph]].

\bibitem{Alvarez:2011hi}
  E.~Alvarez, L.~Da Rold, J.~I.~S.~Vietto and A.~Szynkman,
  JHEP {\bf 1109}, 007 (2011)
  [arXiv:1107.1473 [hep-ph]].

\bibitem{Chivukula:2010fk}
  R.~S.~Chivukula, E.~H.~Simmons and C.~P.~Yuan,
  Phys.\ Rev.\  D {\bf 82} (2010) 094009
  [arXiv:1007.0260 [hep-ph]].

\bibitem{Bai:2010dj}
  Y.~Bai, B.~A.~Dobrescu,
  JHEP {\bf 1107}, 100 (2011).
  [arXiv:1012.5814 [hep-ph]].

\bibitem{afbCMS}
CMS Collaboration, note CMS PAS TOP-10-010.

\bibitem{Krohn:2011tw}
  D.~Krohn, T.~Liu, J.~Shelton and L.~T.~Wang,
  arXiv:1105.3743 [hep-ph].

\bibitem{Collaboration:2011rq}
  V. Abazov {\it et al.} [D0 Collaboration],
  arXiv:1107.4995 [hep-ex].

\bibitem{afbCMS2}
CMS Collaboration, note CMS PAS TOP-11-014.

\end{thebibliography}
\end{document}